\documentclass[prl,twocolumn,aps]{revtex4}
\usepackage{verbatim}
\usepackage{graphicx}
\normalfont
\begin{document}


\title{Long nuclear spin decay times controlled by optical pumping in individual quantum dots}

\author{M. N. Makhonin$^{1}$, A. I. Tartakovskii$^{1}$, I. Drouzas$^{1}$, A. Van'kov$^{1}$, T. Wright$^{1}$, J. Skiba-Szymanska$^{1}$, A. Russell$^{2}$, V. I. Fal'ko$^{2}$, M. S. Skolnick$^{1}$, H.-Y. Liu$^{3}$, M. Hopkinson$^{3}$}

\address{$^{1}$ Department of Physics and Astronomy, University of Sheffield, S3 7RH,UK \\ $^{2}$ Department of Physics, University of Lancaster, Lancaster LA1 4YB, UK\\ $^{3}$ Department of Electronic and Electrical Engineering, University of Sheffield, Sheffield S1 3JD, UK}

\date{\today}

\begin{abstract}
{Nuclear polarization dynamics are measured in the nuclear spin bi-stability regime in a single optically pumped InGaAs/GaAs quantum dot. The controlling role of nuclear spin diffusion from the dot into the surrounding material is revealed in pump-probe measurements of the non-linear nuclear spin dynamics. We measure nuclear spin decay times in the range 0.2-5 sec, strongly dependent on the optical pumping time. The long nuclear spin decay arises from polarization of the material surrounding the dot by spin diffusion for long ($>$5sec) pumping times. The time-resolved methods allow the detection of the unstable nuclear polarization state in the bi-stability regime otherwise undetectable in cw experiments.\\} 
\end{abstract}
\maketitle

The hyperfine interaction between the magnetic moments of the electron and nuclear spins \cite{Overhauser} has been shown to limit the electron spin life-time  \cite{Erlingsson,Merkulov,Braun1,Koppens1,Akimov1,Oulton} and coherence \cite{Khaetskii,Petta,Koppens2} in semiconductor nano-structures. Optical excitation \cite{Gammon,Eble,Lai} and transport \cite{Koppens1,Petta,Koppens2,Ono,Yusa} of spin-polarized electrons in semiconductor quantum dots (QD) have been found to lead to dynamic nuclear polarization: Overhauser magnetic fields up to a few Tesla have been detected in nano-structures \cite{Braun2,Maletinsky1,Tartakovskii} leading to strong modifications of the confined electron energy spectrum \cite{Gammon,Eble,Lai,Braun2,Maletinsky1,Tartakovskii}. The control of the nuclear spin has been identified as one of the prerequisites for the coherent manipulation of the electron spin in semiconductor nano-structures \cite{Klauser}.  
In this context, extended nuclear spin polarization life-times in a dot are desirable, and, as we show here, can be achieved by increasing the optical pumping time, during which not only the dot, but also the surrounding material becomes polarized. 

In this work we measure nuclear spin dynamics in the optically induced bi-stability regime in InGaAs/GaAs self-assembled quantum dots in external magnetic fields $B_{ext}$ of 1-3T \cite{Braun2,Maletinsky1,Tartakovskii}. We develop a sensitive time-resolved all-optical method based on the non-linear behavior of the nuclear spin polarization in the bi-stability regime and measure the rise and decay time of the nuclear polarization in a single dot. Using this method, we find that the optical pumping time required to reach a nuclear spin polarization of $>30\%$ in $B_{ext}=2$T can be as short as 100 ms. Working in a pump-probe mode, where the dot is free of electrons and holes during the ''dark'' pump-probe delay time, we find that the nuclear polarization decay in a single dot is strongly dependent on the pumping time: it can vary from ~0.2 sec for 100ms pumping to ~5 sec for $>3$sec pumping. This observation indicates the importance of nuclear spin pumping outside the dot, which arises due to nuclear spin diffusion from the dot into the surrounding matrix.  Finally, the pump-probe data allow us to reconstruct the whole nuclear spin bi-stability curve, including the unstable nuclear polarization state, which is otherwise undetectable in cw experiments.

We present results for QDs grown in the intrinsic region of an n-type Schottky GaAs/AlGaAs diode. Detailed description of a very similar sample architecture can be found elsewhere \cite{Kolodka}. Low temperature ($T=15$K) photoluminescence (PL) experiments have been performed on individual QDs in the bias regime where the dominant electron (e) and hole (h) population configurations on the dot are ehh ($X^+$) and eh ($X^0$). The sample was mounted in a magnetic field cryostat in the Faraday geometry. 

It has been demonstrated recently that when an InGaAs dot in an external magnetic field of 1-3T is pumped with $\sigma^-$ polarized light, nuclear spin bi-stability is observed (see theoretical curve in Fig.1a) \cite{Braun2,Maletinsky1,Tartakovskii}. In a cw optical experiment, only the red part of the curve, representing the steady-state nuclear polarization, $S_N$, is observed, while the unstable part (blue in Fig.1a) is not detected. As the power is increased above $P_2$ an abrupt transition to the high $S_N$ branch occurs. If then the power is reduced, a threshold-like transition to the low $S_N$ branch is observed at $P_1$. For $P_1<P<P_2$, the bi-stability regime is observed, where two stable magnitudes of $S_N$ are possible. The optically pumped nuclear polarization is evidenced in single-dot PL spectra through the occurrence of significant Overhauser fields ($B_N \propto S_N$) leading to modification of the exciton Zeeman splitting, $E_{xZ}$  \cite{Gammon,Eble,Lai,Braun2,Maletinsky1,Tartakovskii}. This change of $E_{xZ}$ (measured for a single dot) is due to the  modification of the electron Zeeman splitting $E_{eZ}=|g_e|\mu_B(B_{ext}-B_N)$, when $\sigma^-$ polarized excitation is used. Here $g_e$ is the electron g-factor and $\mu_B$ is Bohr magneton.
The switching behavior in Fig.1a occurs due to the feedback of $S_N$ on the nuclear spin pumping rate \cite{Braun2,Maletinsky1,Tartakovskii} via the $B_N$ term in $E_{eZ}$ \cite{hole}. 

Examples of theoretical curves calculated in \cite{Russell} and describing the nuclear spin dynamics in the optically induced bi-stability regime are plotted in Fig.1b and show the time derivative $dS_N/dt$ as a function of $S_N$. The theory predicts only one steady-state $S_N$ solution for high and low power optical pumping of the dot (see gray curves). For intermediate pumping powers, the system enters a three-solution regime, where the one in the middle (marked $S_{unst}$) is unstable, whereas the other two (marked $S_1$ and $S_2$) are stable (see blue curve).

\begin{figure}
\centering
\includegraphics[height=7cm]{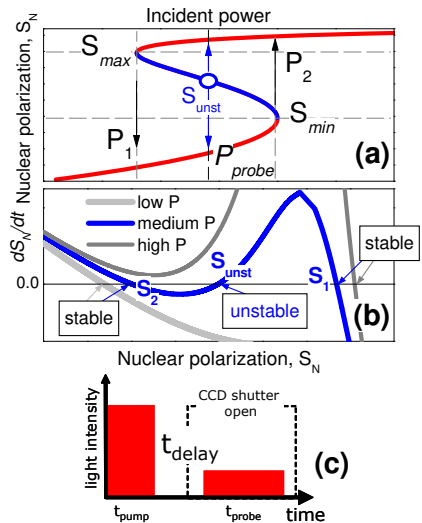}
\caption{(color online) (a) A sketch of a nuclear spin bi-stability curve: nuclear spin polarization in a single dot as a function of optical power. Red (blue) parts of the curve correspond to the stable (unstable) nuclear polarizations on the dot. $P_2$ ($P_1$) mark the threshold-like transitions to the high (low) nuclear polarization branch. An unstable polarization state for an excitation power $P_{probe}$ is marked $S_{unst}$ with the arrow up (down) pointing towards the steady-state nuclear polarizations when the initial polarization is above (below) $S_{unst}$. $S_{max}$ and $S_{min}$ mark the range where the evolution of $S_N$ is measured in our time-resolved experiments. (b) Calculation of the nuclear spin dynamics on the dot: $dS_N/dt(S_N)$. The low and high excitation power curves are shown in gray. For the medium power curve (blue), the stable, $S_1$ and $S_2$, and unstable, $S_{unst}$, states are marked. (c) Diagram of the pump-probe experiment, depicting the light pulse sequence employed.}
\label{fig1} 
\end{figure}

We now propose a sensitive pump-probe method to measure the nuclear spin evolution in a single QD based on the non-linear polarization dynamics in the bi-stability regime.
In our experiments a strong pump pulse excites a single dot for a time $t_{pump}$ (see diagram in Fig.1c). The PL signal is then measured using a CCD and a double monochromator during  excitation with a probe pulse of duration $t_{probe}$ delayed with respect to the pump by $t_{delay}$.  Both pump and probe are $\sigma^-$-polarized and excite the sample into the low energy tail of the wetting layer states, about 120 meV above the lowest QD exciton state. The pump power corresponds to the high power regime observed above $P_2$ in Fig.1a. The probe power is in the bi-stability regime, $P_1<P_{probe}<P_2$. We employ $t_{probe}=8$ sec in order to reach the steady-state nuclear polarization during the probe excitation. During the delay time $t_{delay}$ the dots are strictly empty with neither optically nor electrically generated charges present, so that the evolution of the nuclear spin polarization is not affected by the e-h dynamics during the ''dark time'' (in contrast to \cite{Maletinsky2}).

The pump-probe method is based on the following property of the nuclear spin dynamics in the bi-stability regime: Depending on the initial nuclear polarization, $S_N(t_{delay})$, when the probe excitation is switched on, the final steady-state can occur either at $S_N=S_1$ or $S_N=S_2$ in Fig.1b. If a probe pulse with the power marked $P_{probe}$ in Fig.1a is switched on at a moment when $S_N(t_{delay})<S_{unst}(P_{Probe})$, the nuclear spin will  decay to the low polarization branch (see Fig.1a) as follows from $dS_N/dt<0$ in Fig.1b. In contrast, a high $S_N$ will be observed if $S_N(t_{delay})>S_{unst}(P_{probe})$. In the experiment, for each $t_{delay}$ we find the {\it smallest} $P_{probe}$ for which the steady-state nuclear spin is driven to the high nuclear polarization state (marked $S_1$ in Fig.1b). The  $P_{probe}(t_{delay})$ dependence then reflects the decay with time of the nuclear spin excited by the pump pulse.  Note that the method we employ implies that $S_N$ very close to $S_{unst}$ is probed for each $P_{probe}$. 

The switching from $S_2$ to $S_1$ is observed in PL as a clear reduction of the QD exciton Zeeman splitting by 50-80 $\mu$eV. When no switching takes place ($S_N=S_2$), a large exciton Zeeman splitting, $E_{xZ}\approx |g_e+g_h|\mu_B B_{ext}$ is observed due to the weak contribution of $B_N$, whereas a strong modification of $E_{xZ}\approx |g_e+g_h|\mu_B B_{ext} - |g_e|\mu_B B_N$ is found for $S_N=S_1$. 

Fig.2a shows several examples of dynamics curves measured for a single InGaAs dot employing the method explained above. The curves are measured at $B_{ext}=2$T and correspond to different durations of the pump pulse: 0.3, 0.75 and 10 sec (shown with squares, triangles and circles, respectively). All curves exhibit a common dependence: the probe power required to drive the nuclear spin in the dot to the high polarization branch increases with $t_{delay}$. This corresponds to the gradual decay during $t_{delay}$ of the nuclear polarization excited by the pump. As was shown in our previous work, the initial nuclear spin polarization generated by the pump, $S_0$, can be determined from measurements of the Overhauser shift in PL. 

\begin{figure}
\centering
\includegraphics[height=8cm]{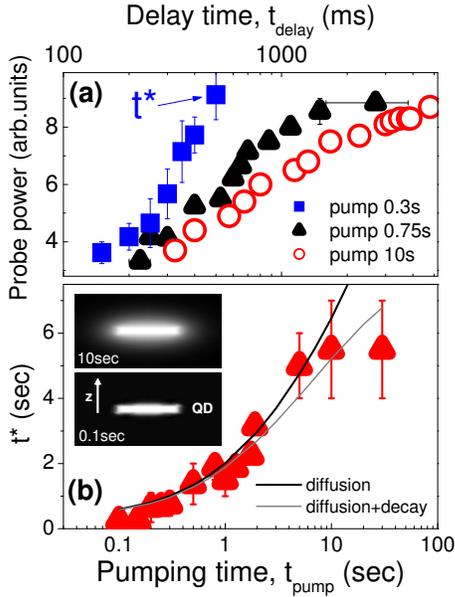}
\caption{(color online) (a) Pump-probe curves, $P_{probe}(t_{delay})$, obtained for pump durations, $t_{pump}$, of 0.3 (squares), 0.75 (triangles) and 10 (circles) sec and probe duration $t_{probe}=8$sec. The last point in each curve corresponds to the characteristic decay time $t^*$ as shown for the curve with $t_{pump}=0.3$sec. $t^*$ is plotted in (b) as a function of $t_{pump}$ (triangles). The contour-plots in (b) show calculated 2D cross-sections of the nuclear spin distribution in a dot and the surrounding material for $t_{pump}=0.1$ (bottom) and 10 (top) sec. The black line shows results of simulations using Eq.\ref{diffusion} with $D_{QD}=5\cdot10^{-14}$cm$^{-2}$/sec. The curve obtained with additional polarization decay rate of 10$^{-2}$ sec$^{-1}$ is shown in gray.}
\label{fig2} 
\end{figure}

The data points for the longest $t_{delay}$ on the decay curves in Fig.2a were measured for $P_{probe}\approx  0.9P_2$ (see $P_2$ label in Fig.1a). These correspond to the maximum delay time and lowest $S_N$ that can be measured reliably in this experiment. For higher $P_{probe}$, the nuclear spin on the dot is driven to the high $S_N$ state even without the pump excitation. As evidenced from the measurement of the Overhauser shift, at $B_{ext}=2$T this cut-off occurs at $S_{min}\approx 0.11$ (see Fig.1a and Fig.3), below which the remaining nuclear polarization generated by the pump produces no change of the polarization generated by the probe. In Fig.2a we denote the time where the curves reach this cut-off polarization level as $t^*$. As seen from Fig.2a, $t^*$ increases markedly with $t_{pump}$, indicating slower nuclear polarization decay for longer pumping times.

This dependence is summarized in Fig.2b. We find that for $t_{pump}<0.1$sec nuclear spin switching to the high polarization state is not possible.  For $t_{pump}$ varied from 0.1 to 2 sec, the function $t^*(t_{pump})$ exhibits a marked increase from 0.25 to 3 sec.  $t^*(t_{pump})$ then saturates at $\approx$5 sec for $t_{pump}\geq$5 sec. Irrespective of the pumping time, the degree of nuclear polarization inside the dot reaches the same maximum magnitude of $\approx 0.32$ corresponding to an Overhauser shift of $\approx 88\mu$eV. This observation together with the strong dependence $t^*(t_{pump})$ in Fig.2b can only arise if the rate of spin depolarization on the dot is governed by the finite nuclear spin polarization outside the dot. Note that there are no charge carriers in the QD during the ''dark'' delay times which may affect nuclear polarization as in \cite{Maletinsky2}. The polarization of nuclei directly via optical excitation in the wetting layer is unlikely since the carriers relax to the lowest QD states during several ps after excitation~\cite{Zibik}. The most likely scenario for the polarization outside the dot is diffusion of the nuclear spin initially excited on the dot into the bulk during the pumping time \cite{Paget}.  

The evidence for diffusion is supported by the results of our modeling. The contour-plots in Fig.2b show 2D cross-sections of the nuclear polarization occurring in the dot and surrounding bulk during the pump excitation with a short (bottom) and long (top) $t_{pump}$.  These figures are calculated by numerically solving a standard 3D diffusion equation 
\begin{equation}
dS_N(r,t)/dt=D_{QD}\Delta S_N(r,t),
\label{diffusion}
\end{equation}
with a condition of time-independent polarization in the dot volume, approximated by a disk of a low aspect ratio. Here $D_{QD}$ is the nuclear spin diffusion coefficient.  As seen in the upper contour plot in Fig.2b, for short $t_{pump}$, the high polarization volume is mainly confined to the dot itself. In this case, after the pump is switched off, strong leakage of the nuclear spin occurs through the large dot surfaces on the top and bottom of the dot, leading to a fast 1D-like spin diffusion and rapid decay of the nuclear spin on the dot. A much larger volume becomes polarized if longer pumping is employed due to spin diffusion from the highly polarized dot into the surrounding matrix, as shown in the lower contour plot of Fig.2b, with consequential slowing down of the diffusion rate and hence nuclear spin decay rate.

In order to model the $t^*(t_{pump})$ dependence in Fig.2b, we calculate the decay of $S_N$ in the dot to the cut-off level $S_{min}$, using as the initial spatial distribution of $S_N$ the polarization established in and around the dot during the pumping time. The decay time obtained in such a way corresponds to the experimentally determined $t^*(t_{pump})$ function and is plotted in Fig.2b (black line). We used a 4x20 nm disk-shaped dot \cite{TEM} and obtained the best fit for $D_{QD}=5 \cdot 10^{-14}$cm$^{-2}$/s consistent with previously reported values for bulk GaAs ($D_{GaAs}\approx 10^{-13}$cm$^2$/s \cite{Paget}). Very good agreement with experiment is achieved in the range $t_{pump}\leq 10$ sec. The discrepancy between theory and experiment observed for $t_{pump}>10$ sec may originate from additional nuclear spin decay processes such as spin-lattice relaxation with characteristic decay times of the order of $10^2\div10^3$ sec \cite{spinlattice} which will lead to the observed saturation of $t^*(t_{pump})$ at long $t_{pump}$. A better fit to the experiment at long $t_{pump}$ is obtained with an additional spin decay term introduced to Eq.~\ref{diffusion} (gray line in Fig.2b).

\begin{figure}
\centering
\includegraphics[height=4.5cm]{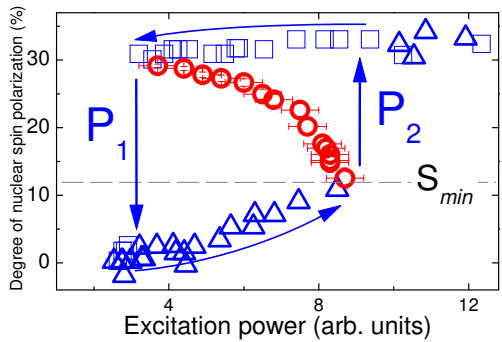}
\caption{(color online) Degree of nuclear polarization on the dot as a function of the optical pumping power. Triangles (squares) show results for the power scan up (down). Thresholds are marked similarly to Fig.1a as $P_2$ and $P_1$. The data representing the unstable nuclear spin state deduced from the data in Fig.2a are plotted with circles.}
\label{fig3} 
\end{figure}

Our calculations show that in the slow polarization decay regime the decay of the nuclear spin can be approximated with a time-independent rate. Using this assumption we estimate the slowest nuclear spin depolarization rate in our dot, $w^{min}_{dep}$, achieved for $t_{pump}=10$ sec. The maximum and minimum (cut-off) magnitudes of $S_N$ probed in the dynamics experiment are determined from a measurement of the bi-stability curve using single beam excitation (Fig.3). The threshold powers where transitions to high and low $S_N$ take place are labeled as in Fig.1a as $P_2$ and $P_1$, respectively.  From the magnitude of the Overhauser shift in the single-beam measurement we estimate $S_N\approx $0.11 and 0.30 just before the thresholds at $P_2$ and $P_1$, respectively.  For all curves in Fig.2a the data points measured at the shortest and longest $t_{delay}$ correspond to excitation with $P_{probe}\approx P_1$ and $P_2$, respectively. We now use $S_N(t_{delay})=S_0exp(-t_{delay}w^{min}_{dep})$ and consider the curve with $t_{pump}=10$sec (circles in Fig.2a). Substituting the magnitudes of $S_N$ for $P_2$ and $P_1$, we obtain $w^{min}_{dep}= 0.19\pm0.02$ sec$^{-1}$. Slow decay rates are thus achieved by prolonged optical pumping, even though the dot and the surrounding matrix are composed of similar material, where rather efficient spin diffusion and hence fast polarization decays would normally be expected. 

As was described above, each data point $P_{probe}(t_{delay})$ corresponds to the nuclear polarization $S_N(t_{delay})\approx S_{unst}(P_{probe})$. By applying the approximation of time-independent polarization decay for long $t_{pump}$, we can now re-plot $P_{probe}(t_{delay})$ as $S_N(P_{probe})$ using the depolarization rate $w^{min}_{dep}$ found above. The function $S_N(P_{probe})$ obtained for $S_0=0.32$ and $w^{min}_{dep}= 0.19$ sec$^{-1}$ is plotted in Fig.3 as the circles, and with high accuracy corresponds to $S_{unst}$ as a function of optical power (as shown in Fig.1a,b).

In conclusion, in this work we measure nuclear spin dynamics in a single InGaAs/GaAs quantum dot in the regime of optically induced nuclear spin bi-stability in magnetic field of 1-3T. We develop a new method sensitive to weak variations of the nuclear spin on the dot that is based on the strongly non-linear nuclear spin dynamics in the bi-stability regime. Using this method we find that nuclear spin excited on the dot diffuses into the surrounding dot matrix. This is in contrast to recent studies of the nuclear spin dynamics in similar single InGaAs dots at $B_{ext}\leq0.22$T \cite{Maletinsky2}. The reason for weak diffusion effects for low fields is currently unclear. As we show the nuclear spin polarization decay can be suppressed by polarizing the surrounding matrix, a process which can be controlled by the optical pumping time.  

We thank M. Gordovskyy for fruitful discussions. This work has been supported by the Sheffield EPSRC Programme grant GR/S76076, the EPSRC IRC for Quantum Information Processing, ESF-EPSRC network EP/D062918, by the Royal Society, and by the Lancaster-EPSRC Portfolio Partnership EP/C511743. AIT was supported by the EPSRC Advanced Research Fellowship EP/C54563X/1 and research grant EP/C545648/1.


\begin{thebibliography}{99}
\bibitem{hole} The hole-nuclear spin interaction is negligible due to the p-type of the hole wave-function.
\end{thebibliography}
\end{document}